\documentclass[onecolumn,showpacs,preprintnumbers,amsmath,amssymb,superscriptaddress]{revtex4-2}

\usepackage{epsf}
\usepackage{graphicx}
\usepackage{sidecap}
 \usepackage{soul}
 \usepackage{array}
 \usepackage{amsmath}
 \usepackage{amssymb}
 \usepackage{dcolumn}
 \usepackage{epstopdf}
 \usepackage{bm}
 \usepackage{amsmath}
 
\usepackage{gensymb}
\usepackage{color}
\usepackage{hyperref}
\usepackage{soul}
\sethlcolor{green}
\usepackage{bbding}
\hypersetup{
    colorlinks=true,
    citecolor=red,
    linkcolor=red,
    filecolor=blue,   
    urlcolor=blue,
}
 
\def \beq {\begin{equation}}
\def \eeq {\end{equation}}
\begin{document}
 
\title{{Spectroscopic evidence of flat bands in breathing kagome semiconductor  ${\mathrm{\textbf{Nb}}}_{3}{\mathrm{\textbf{I}}}_{8}$ }}
\author{Sabin~Regmi} \affiliation{Department of Physics, University of Central Florida, Orlando, Florida 32816, USA}
 \author{Tharindu Fernando} \affiliation{Department of Physics, University of Washington, Seattle, Washington 98195, USA}
 \author{Yuzhou Zhao}\affiliation{Department of Physics, University of Washington, Seattle, Washington 98195, USA}
 \author{Anup~Pradhan~Sakhya}\affiliation{Department of Physics, University of Central Florida, Orlando, Florida 32816, USA}
 \author{Gyanendra~Dhakal}\affiliation{Department of Physics, University of Central Florida, Orlando, Florida 32816, USA}
 \author{Iftakhar Bin Elius} \affiliation{Department of Physics, University of Central Florida, Orlando, Florida 32816, USA}
   \author{Hector Vazquez}\affiliation{Department of Physics, University of Central Florida, Orlando, Florida 32816, USA}
   \author{Jonathan D Denlinger} \affiliation{Lawrence Berkeley National Laboratory, Berkeley, California 94720, USA}
  \author{Jihui Yang} \affiliation{Department of Materials Science and Engineering, University of Washington, Seattle, Washington 98195, USA}
  \author{Jiun-Haw Chu} \affiliation{Department of Physics, University of Washington, Seattle, Washington 98195, USA}
  \author{Xiaodong Xu}\affiliation{Department of Physics, University of Washington, Seattle, Washington 98195, USA}
\author{Ting Cao} \affiliation{Department of Materials Science and Engineering, University of Washington, Seattle, Washington 98195, USA}
 \author{Madhab Neupane} \thanks{Corresponding author: \href{mailto:madhab.neupane@ucf.edu}{madhab.neupane@ucf.edu}}\affiliation{Department of Physics, University of Central Florida, Orlando, Florida 32816, USA}
\begin{abstract}
\textit{Abstract - }Kagome materials have become solid grounds to study the interplay among geometry, topology, correlation, and magnetism. Recently, niobium halide semiconductors $\mathrm{Nb}_{3}X_{8} (X~=~\mathrm{Cl},~\mathrm{Br},~\mathrm{I})$ have been predicted to be two-dimensional magnets and these materials are also interesting for their breathing kagome geometry. However, experimental electronic structure studies of these promising materials are still lacking. Here, we report the spectroscopic evidence of flat and weakly dispersing bands in breathing-kagome semiconductor $\mathrm{Nb}_{3}\mathrm{I}_{8}$ around 500 meV binding energy, which is well supported by our first-principles calculations. These bands originate from the breathing kagome lattice of niobium atoms and have niobium \textit{d}-orbital character. They  are found to be sensitive to polarization of the incident photon beam. Our study provides insight into the electronic structure and flat band topology in an exfoliable kagome semiconductor, thereby providing an important platform to understand the interaction of geometry and electron correlations in two-dimensional materials.
\end{abstract}
\maketitle

\noindent{\textit{Introduction - }}
Owing to the possibility of exploring the interplay among some or all of parameters such as geometry, topology, electronic correlations and magnetism, quantum materials with kagome lattice have recently been attracting a bulk of research studies \cite{Kida, Han, Nakatsuji, Fe3Sn2_Lin, Yin1, Yin2, Yang, Tb166, Lin, Ghimire}. Comprised of corner sharing triangles forming a hexagon within, a kagome lattice may feature relativistic Dirac band crossing at the corner of the Brillouin zone (BZ) \cite{Dirac}. Because of the structural geometry, the electrons are trapped within the hexagon. This geometry-driven self-localization of the electrons means that flat bands may be present in the electronic structure of the kagome materials and give rise to strong electronic correlations \cite{Fe3Sn2_Lin, CoSn_1, CoSn_2, FeSn, Y166}. By introducing spin-orbit coupling and magnetism that breaks the time-reversal symmetry, the Dirac band crossing at the corner of the BZ can be gapped out leading to intrinsic Chern quantum phases \cite{Tang, Xu, Tb166, Fe3Sn2_2}. The band gap may be opened also when the alternating triangles in the kagome lattice have different bond lengths resulting a different geometry than the conventional kagome (in which all the triangles are identical) - called the breathing kagome \cite{Bolens}- leading to the possibility of higher order topological phase \cite{Ezawa}. The intrinsically topological flat bands, however, may be robust to this geometrical perturbation \cite{Bolens}. The difference in the size of the alternate triangles in the breathing kagome geometry may result in local electric dipole moment leading to ferroelectric order \cite{YLi}.\\

\begin{figure*} [ht!]
\includegraphics[width=0.9\textwidth]{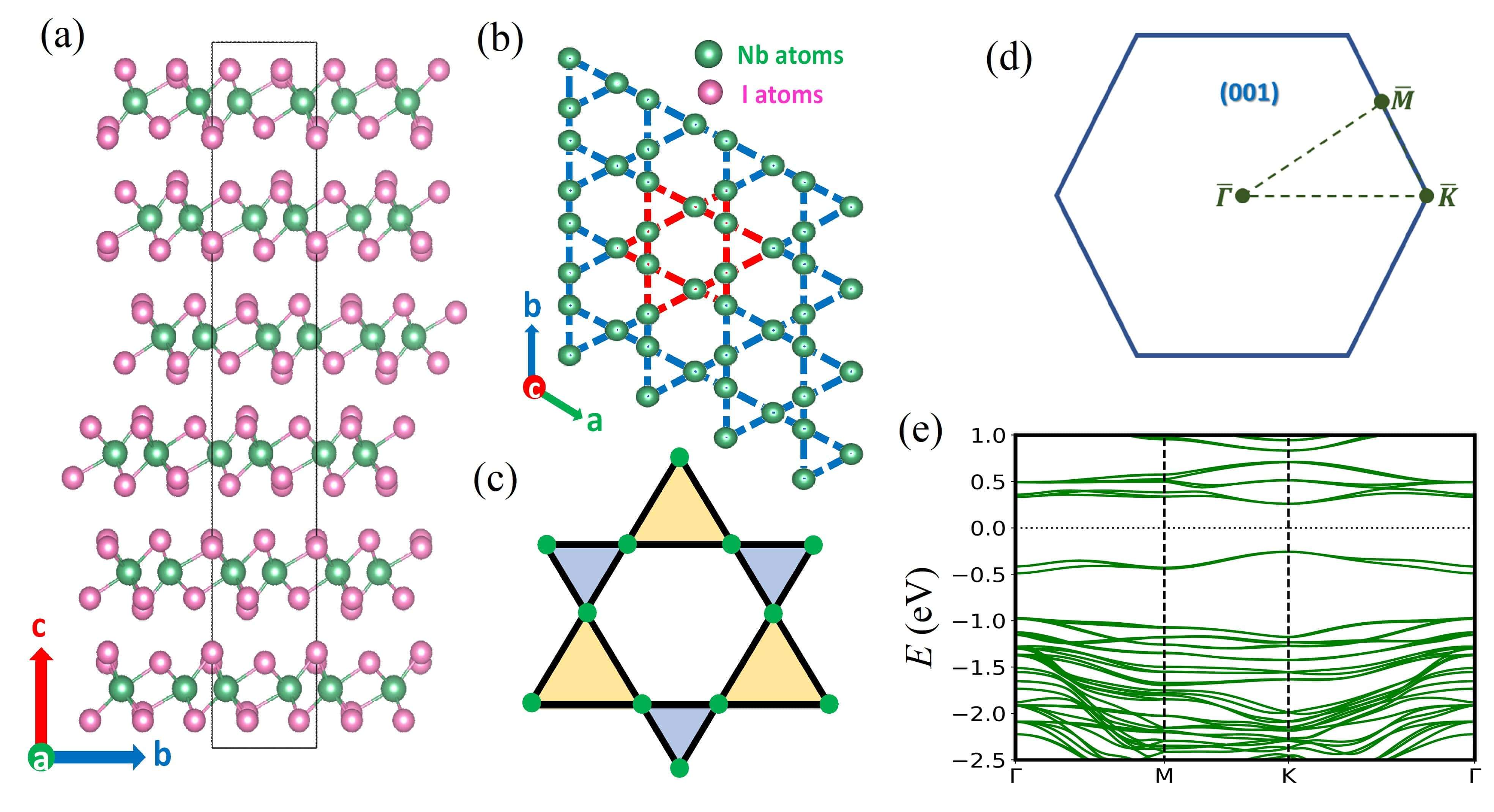}
\caption{\textbf{Crystal structure and bulk band calculation of ${\mathrm{\textbf{Nb}}}_{3}{\mathrm{\textbf{I}}}_{8}$.} \textbf{a} Crystal structure of ${\mathrm{Nb}}_{3}{\mathrm{I}}_{8}$, where the green and magenta spheres represent Nb and I atoms, respectively. \textbf{b} Nb atoms forming a breathing kagome lattice. An schematic of breathing kagome lattice is presented in \textbf{c}. The alternating traingles are not identical giving rise to breathing lattice. \textbf{d} Projected hexagonal BZ on the (001) surface showing the high-symmtery points and directions. \textbf{e} Calculated band structure for bulk ${\mathrm{Nb}}_{3}{\mathrm{I}}_{8}$.} 
\label{F1}
\end{figure*}

Angle-resolved photoemission spectroscopy (ARPES) \cite{ARPES1, ARPES2, ARPES3} has served as a valuable technique allowing direct visualization of  the band structure in quantum materials. In recent times, it has been successful in revealing the electronic structure of various kagome materials. Topological Dirac bands and/or flat bands in the electronic structures of  $\mathrm{Fe_3Sn_2}$ \cite{Fe3Sn2_Lin, Fe3Sn2_2, Tanaka}, $\mathrm{FeSn}$ \cite{FeSn}, $\mathrm{CoSn}$ \cite{CoSn_1, CoSn_2}, $\mathrm{\textit{R}\textit{T}_6Sn_6}$($\mathrm{\textit{R} = ~Tb, Y,~Er,~Ho,~Gd}$ and $\mathrm{\textit{T}~=~Mn,~V )}$ \cite{Tb166, Y166, Er166, Ho166, RV166}, $\mathrm{CsV_3Sb_5}$ \cite{Ortiz} have been observed  by utilizing ARPES. A point to note is that most of these materials are metals with conventional kagome lattice, $\mathrm{Fe_3Sn_2}$ with breathing kagome lattice reported in a recent ARPES study suggests a possibility of magnetic Weyl semimetallic phase in this material \cite{Tanaka}.  ARPES study of materials with the breathing kagome lattice is still limited. The ${\mathrm{Nb}}_{3}{\mathrm{\textit{X}}}_{8}$ family of materials, where $\mathrm{\textit{X}~=~Cl,~ Br, ~I}$ \cite{Magonov, JiangNb3X8}, provides material platforms to study the electronic structure arising from the breathing kagome geometry. Importantly, these materials have two-dimensional (2D) crystalline structure with very weak interlayer van der Waals interaction and hence can be thinned down to 2D limit via mechanical exfoliation \cite{Kim, Oh, Yoon}. They are semiconducting with a moderate band gap \cite{Oh}, which is suitable for electronic applications. In the monolayer limit, they are predicted to host  ferromagnetic (FM) order \cite{JiangNb3X8, Conte, Conte2}, allowing a great platform to study the interplay among geometry, electronic correlations, and magnetism in 2D. Monolayer ${\mathrm{Nb}}_{3}{\mathrm{I}}_{8}$ is predicted to exhibit large spontaneous valley polarization making it suitable for valleytronics applications \cite{RuiPeng}. Systematic studies of the electronic structure of these semiconducting van der Waals breathing kagome materials are worthwhile, however, are lacking.  \\

In this article, via ARPES measurements, we report a systematic electronic structure study of breathing kagome semiconductor ${\mathrm{Nb}}_{3}{\mathrm{I}}_{8}$ assisted by density-functional theory (DFT) computations. Our experimental data is consistent with the semiconducting nature of ${\mathrm{Nb}}_{3}{\mathrm{I}}_{8}$ and reveals the presence of flat and weakly dispersing bands, that arise from the Nb-breathing kagome, in the electronic band structure of this material. These bands are observed to be sensitive to the polarization of the incident photon beam. The experimental observations are well reproduced by theoretical calculations. Our study provides a layered 2D material platform with flat bands thereby giving an opportunity to study the interaction among symmetry, geometry, electronic correlations and potentially magnetism given the prediction of FM monolayer. \\

\begin{figure*} [ht!]
\includegraphics[width=1\textwidth]{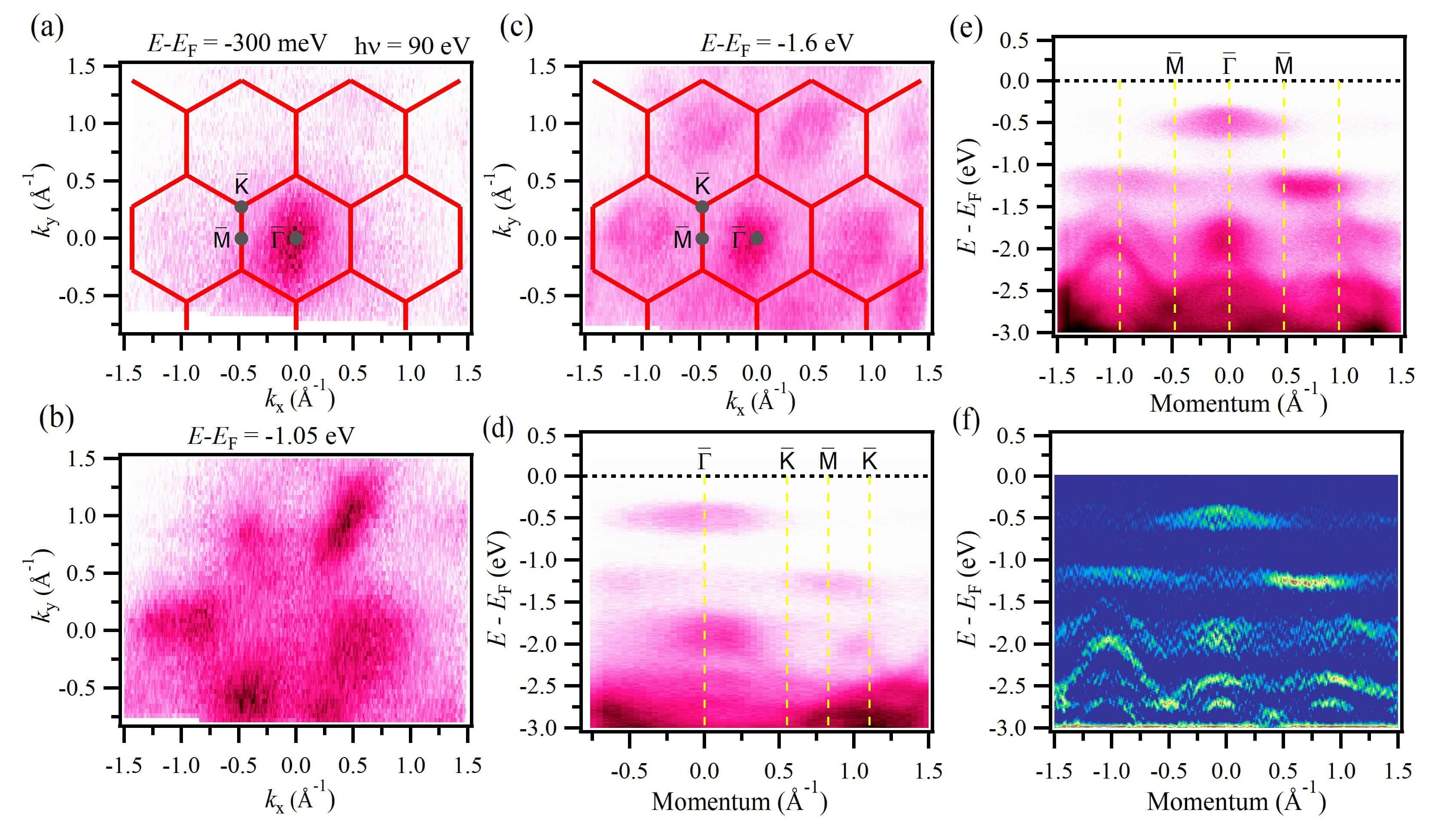}
\caption{\textbf{ARPES measured energy contours (ECs) and band dispersions.} \textbf{a-c} ECs at the binding energies of 300 meV, 1.05 eV, and 1.6 eV, respectively. Red hexagons represent the BZ with high-symmetry points marked and labeled. \textbf{d-e} Band dispersions along different high-symmetry directions as shown in the plots. \textbf{f} Second derivative plot of e. Data were collected at the ALS beamline 4.0.3 at a temperature of 260 K using photon source of mono-energy 90 eV. }
\label{F2}
\end{figure*}

\noindent\textit{Methods - }
High-quality single crystals of ${\mathrm{Nb}}_{3}{\mathrm{I}}_{8}$ were synthesized by utilizing chemical vapor transport method (modified from a previously reported method \cite{Kim}) using Iodine as a transport agent (see Supplementary note 1).  The crystal structure was examined using X-ray diffraction and the chemical compositions of the obtained crystals were verified using scanning electron microscopy and energy dispersive X-ray spectroscopy analysis (See Supplementary Fig. 2).\\ 

Synchrotron based ARPES measurements were performed at the Advanced Light Source (ALS) beamline 4.0.3 equipped with R8000 hemispherical analyzer. In order to obtain high-quality and fresh surfaces required for ARPES experiments, single crystals were  cleaved \textit{in situ} under ultra high vacuum with pressure maintained in the order of $\mathrm{10^{-11}}$ Torr.  The measurements were carried out at a temperature of 260 K to evade charging effects. This leads to thermal broadening of the bands in the ARPES spectra. LH and LV polarized photon sources in the range $\mathrm{30 - 124~eV}$ were used during the measurements. For experimental geometry, see Supplementary Note 3 and Supplementary Fig. 6.\\

Computational results were obtained from first-principles calculations performed within the framework of DFT \cite{DFT1, DFT2} as implemented in the {\sc Quantum ESPRESSO} package \cite{QE} and Vienna ab initio simulation package ({\sc VASP}) \cite{Kresse1996}. Norm-conserving scalar relativistic ONCVPSP (Optimized Norm-Conserving Vanderbilt Pseudopotential) \cite{ONCVP1, ONCVP2} and Perdew-Burke-Ernzerhof-type exchange-correlational functional \cite{PBE} were used with van der Waals correction included in the D2 formalism \cite{D2}. For more information, see Supplementary Note 2.\\

\noindent\textit{Results - }
${\mathrm{Nb}}_{3}{\mathrm{I}}_{8}$ crystallizes in trigonal crystal system with rhombohedral space group $\mathrm{R\overline{3}m}$ (number 166) with lattice parameters ${a}$ = ${b}$ = $\mathrm{7.60~\AA}$ and ${c}$ = $\mathrm{41.715~\AA}$ \cite{Magonov}. A three-dimensional bulk crystal structure with six layers within a unit cell is shown in Fig. \ref{F1}a where each layers along the crystallographic $\mathrm{c}$-direction are connected through a weak van der Waals bond and within each layer, there is strong covalent bonding. The Nb atoms form a breathing kagome lattice with two distinct Nb-Nb distances in alternate Nb triangles and are sandwiched between the top and the bottom I-atom sheets, each containing two atomic layers (I1, I3 on the top sheet and I2, I4 on the bottom sheet) (see Supplementary Note 1 and Supplementary Fig. 1). The breathing kagome formed by the Nb-atoms can be seen in Fig. \ref{F1}b, where each Nb atom has two long inter-triangular distances and two short intra-triangular distances with its four nearest neighbor Nb-atoms within the kagome plane. A schematic of the breathing lattice is presented in Fig. \ref{F1}c, where the alternate triangles, unlike in a conventional kagome lattice, are not identical. Because of the presence of this special geometry in  ${\mathrm{Nb}}_{3}{\mathrm{I}}_{8}$, flat bands are expected in its electronic structure. In Fig. \ref{F1}d, a projected hexagonal BZ on (001) surface is presented with high-symmetry points and directions. In Fig. \ref{F1}e, we present the calculated band structures for bulk ${\mathrm{Nb}}_{3}{\mathrm{I}}_{8}$ at $k_z$ = 0. According to the calculation, ${\mathrm{Nb}}_{3}{\mathrm{I}}_{8}$ is  semiconducting with a Kohn-Sham band gap (direct) of $\sim \mathrm{520~meV}$ at the $\mathrm{K}$ point. Flat bands and weakly dispersing bands can be seen within 500 meV above and below the mid-gap. Such bands are present around 1 eV below the Fermi level as well. The overall frontier band structures of the bulk seems to be similar to that of monolayer, indicating that the interlayer electronic couplings are weak (Supplementary Fig. 3 for monolayer calculation). \\

\begin{figure*} [ht!]
\includegraphics[width=1\textwidth]{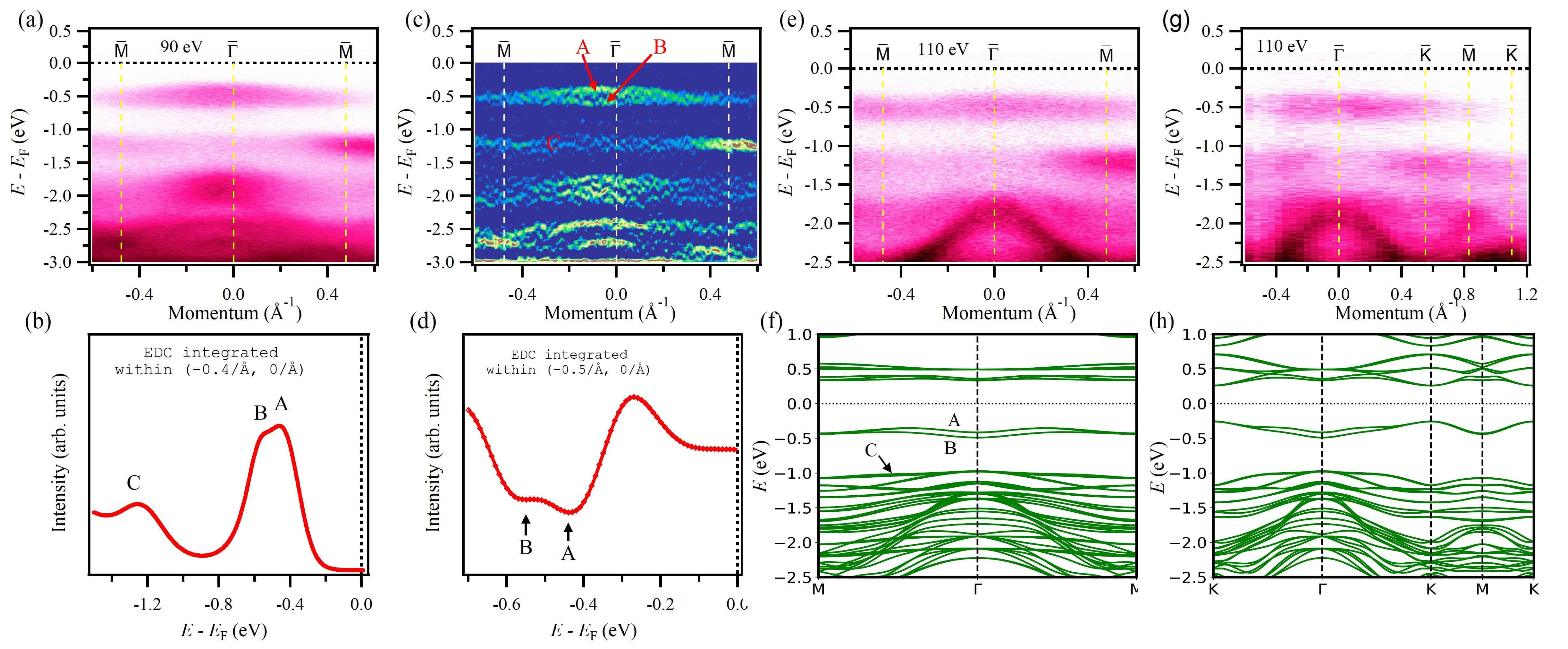}
\caption{\textbf{Observation of flat and weakly dispersing bands.} \textbf{a,b} ARPES measured band dispersions along $\overline{\mathrm{M}} - \overline{\mathrm{\Gamma}} - \overline{\mathrm{M}}$ measured using photon energy of 90 eV and Integrated EDC (Voigt fit, also see Supplementary Note 6 and Supplemenatry Fig. 9) within an momentum window of ($\mathrm{-0.4~/\AA}$ - $\mathrm{0~/\AA}$) in a. \textbf{c,d} Second derivative plot of dispersion map in a  and respective EDC curve taken within momentum window of ($\mathrm{-0.5~/\AA}$ - $\mathrm{0~/\AA}$), respectively. \textbf{e} Experimental band dispersion along $\overline{\mathrm{M}} - \overline{\mathrm{\Gamma}} - \overline{\mathrm{M}}$ direction. \textbf{f} Calculated bands along $\mathrm{M}-\mathrm{\Gamma}-\mathrm{M}$. \textbf{g} Experimental band dispersion along $\overline{\mathrm{\Gamma}} - \overline{\mathrm{K}} - \overline{\mathrm{M}} - \overline{\mathrm{K}}$ direction. \textbf{h} Calculated bands along $\mathrm{\Gamma}-\mathrm{K}-\mathrm{M}-\mathrm{K}$. ARPES data were collected at the ALS beamline 4.0.3 at a temperature of 260 K.}
\label{F3}
\end{figure*}

\begin{figure*} [ht!]
\includegraphics[width=1\textwidth]{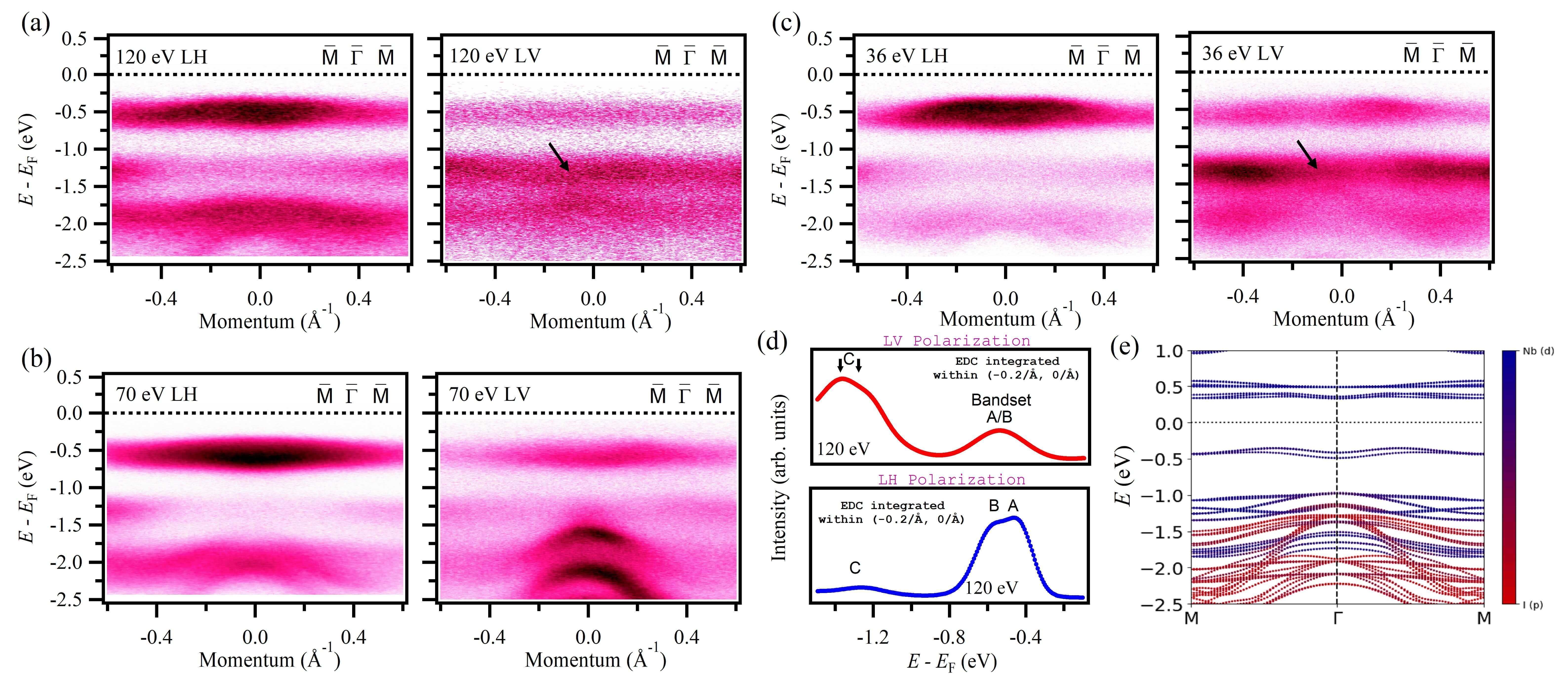}
\caption{\textbf{Polarization dependent band dispersion.} \textbf{a-c} ARPES measured band dispersions along $\overline{\mathrm{M}} - \overline{\mathrm{\Gamma}} - \overline{\mathrm{M}}$ measured using photon energies of 120 eV, 70 eV, and 36 eV, respectively using LH or LV polarization as noted on each plot. \textbf{d} Fits of integrated EDCs (Voigt fits) taken within momentum window of ($\mathrm{-0.2~/\AA}$ - $\mathrm{0~/\AA}$) for 120 eV LH polarized (bottom) and LV polarized (top) data in a [also see Supplementary Fig. 9]. \textbf{e} Calculated orbital-resolved dispersion along ${\mathrm{M}} - {\mathrm{\Gamma}} - {\mathrm{M}}$. ARPES data were collected at the ALS beamline 4.0.3 at a temperature of 260 K.}
\label{F4}
\end{figure*}

In Figs. \ref{F2} - \ref{F4}, we present the electronic structure of ${\mathrm{Nb}}_{3}{\mathrm{I}}_{8}$ measured using ARPES. ${\mathrm{Nb}}_{3}{\mathrm{I}}_{8}$ is a semiconductor and the Fermi level lies within the gap. Therefore, no photoemission signal was observed at the Fermi level. ARPES signal is observed only about 300 meV below the Fermi level [see Fig. \ref{F2}a]. Several pockets can be seen below 1 eV from the Fermi level, as shown in Fig. \ref{F2}b. At a binding energy of about 1.6 eV, a nice hexagonal symmetry is observed in the energy contour [see Fig. \ref{F2}c]. Next, in order to reveal the electronic structure of ${\mathrm{Nb}}_{3}{\mathrm{I}}_{8}$, we present the dispersion maps along various high-symmetry directions in Figs. \ref{F2}d-e. The dispersion maps along $\overline{\mathrm{M}} - \overline{\mathrm{\Gamma}} - \overline{\mathrm{M}}$ and $\overline{\mathrm{\Gamma}} - \overline{\mathrm{K}} - \overline{\mathrm{M}} - \overline{\mathrm{K}}$ show the presence of bands in accordance to the observations in the energy contours. The photoemission intensity is not uniform and specially at around 500 meV, is strongly suppressed out of a single BZ (also seen in Fig. \ref{F2}a), possibly because of matrix element effects.  This can be seen in the second derivative plot of the dispersion along $\overline{\mathrm{M}} - \overline{\mathrm{\Gamma}} - \overline{\mathrm{M}}$, presented in Fig. \ref{F2}f.\\

Next, we proceed to analyze the details of the bands observed along the high-symmetry directions along with a careful comparison with theoretically calculated bands using the DFT approach in Fig. \ref{F3}. Figure \ref{F3}a shows the dispersion map along $\overline{\mathrm{M}} - \overline{\mathrm{\Gamma}} - \overline{\mathrm{M}}$ obtained from ARPES measurement using 90 eV photon energy. Around 500 meV below the Fermi level, there are two bands, one of which is almost flat (named B) and the other is weakly dispersing (named A), which correspond to the peaks in the integrated energy distribution curve (EDC) (Voigt fit) in Fig. \ref{F3}b [also see Supplementary Fig. 9]. It has been predicted that flat and weakly dispersing bands in breathing kagome lattices might be intrinsically robust \cite{Bolens}. Another  nearly flat band (named C) can also be observed below 1.2 eV binding energy. In Fig. \ref{F3}c, we present the second derivative plot of the dispersion map in Fig. \ref{F3}a, where the presence of two bands - one weakly dispersing (A) and the other almost flat (B) - can be clearly seen. Two minima peaks in the integrated EDC taken within ($\mathrm{-0.5~/\AA}$ - $\mathrm{0~/\AA}$) of the second derivative plot (Fig. \ref{F3}d) further confirm the presence of two bands. The minima peaks exactly coincide the maxima peaks in the integrated EDC in Fig. \ref{F3}b. In Figs. \ref{F3}e and \ref{F3}f, we present the experimental band structure along $\overline{\mathrm{M}} - \overline{\mathrm{\Gamma}} - \overline{\mathrm{M}}$ taken with 110 eV photon energy  and calculated band structure using DFT approach, respectively, where we can see an reasonable matching between the two. Within 500 meV of the mid-gap, the calculated data shows a set of almost flat bands (labeled as A and B) that have weak dispersions persisting down to the monolayer limit. Another set of almost flat bands (labeled C) appears around 1 eV below the Fermi level. Below C, DFT calculations show a continuum of band states consistent with the finite intensity observed in the experimental APRES spectra. The flat bands do not seem to vary their dispersion with the change in photon energy regardless of the variation in intensity, which possibly comes from the photoemission matrix element effect. This shows that these bands are of 2D nature, which is expected for 2D materials like ${\mathrm{Nb}}_{3}{\mathrm{I}}_{8}$ [For more variation in photon energy, see Supplementary Note 4 and Supplementary Fig. 7]. The dispersive bands below 1.5 eV binding energy, however, seem to strongly depend on the energy of the incident photon beam indicative of their 3D nature.  Figures \ref{F3}g and \ref{F3}h represent the experimental and calculated band structures along  $\overline{\mathrm{\Gamma}} - \overline{\mathrm{K}} - \overline{\mathrm{M}} - \overline{\mathrm{K}}$, respectively. The calculated result well reproduces the experimental observation. \\

In Fig. \ref{F4}, we present the polarization dependence of the bands observed along the $\overline{\mathrm{M}} - \overline{\mathrm{\Gamma}} - \overline{\mathrm{M}}$ direction. Polarization dependent measurements have been performed at several photon energies including 120 eV, 70 eV, and 36 eV and presented in Figs. \ref{F4}a-c respectively. It can be seen that the bands around 500 meV binding energy have strong intensity in the linear horizontal (LH) polarization [see bottom panel of Fig. \ref{F4}d]. These bands are strongly suppressed in intensity when the incident photon beam is linear vertical (LV) polarized. With LV polarization, the intensity of the bands below 1.2 eV becomes strong [see top panel of Fig. \ref{F4}d]. A weakly dispersing band can also be seen with LV polarized light [shown by arrow in the right panels of Fig. \ref{F4}a and \ref{F4}c], supported by two-peak feature in Fig. \ref{F4}d top panel. In Fig. \ref{F4}e, we present the contribution of $\mathrm{Nb}$ \textit{d} and $\mathrm{I}$ \textit{p} in the electronic structure of ${\mathrm{Nb}}_{3}{\mathrm{I}}_{8}$ along the $\overline{\mathrm{M}} - \overline{\mathrm{\Gamma}} - \overline{\mathrm{M}}$ direction. It can be seen that the bands within 500 meV and just below 1 eV binding energy have strong contribution from Nb \textit{d} and the highly dispersive bands near the center of the BZ and  below 1.5 eV  are strongly dominated by I \textit{p}. The flat bands observed in our experiments, therefore, originate from the breathing kagome lattice of the Nb atoms. We further perform orbital-resolved calculations to study the contribution of each \textit{d} orbitals, namely $d_{zx}$, $d_{x^2-y^2}$, $d_{z^2}$, $d_{yz}$, and $d_{xy}$ [see Supplementary Note 5 and Supplementary Fig. 8 for orbital resolved calculations]. The band sets A and B have strong contribution from $d_{z^2}$ orbitals. On the other hand, the band sets C have significant contribution from all orbitals with majority coming from the $d_{xy}$ orbitals.\\

\noindent\textit{Discussion - }
Because of the breathing kagome geometry of the Nb atoms in ${\mathrm{Nb}}_{3}{\mathrm{I}}_{8}$, the Dirac fermion is not preserved in this system as in the case of the conventional kagome geometry. This might have led to the semiconducting nature of this material as seen in our theoretical calculations and backed by our experimental observation. However, the flat and weakly dispersing bands may remain robust despite the geometric perturbation from kagome to breathing kagome \cite{Bolens}. In our experimental measurements, we observe multiple flat and weakly dispersing bands at various binding energies. Our calculations also show that the frontier flat and weakly dispersing bands of the bulk is similar to those of the monolayer. This makes the choice of \textit{k$_z$} not so significant in the measurements and can be validly compared with the DFT calculations (also see calculations for \textit{k$_z$} = 0 and \textit{k$_z$} = $\pi$ in supplementary Fig. 5, where no significant differences can be seen). Our analysis shows that no band inversion occurs for the bands around $\mathrm{500~meV}$ binding energy indicating non-topological character and quantized non-zero Chern numbers do not exist for the lower bands by rigidly shifting the Fermi level as they cannot be completely isolated in energy (see Supplementary Note 2 and Supplementary Fig. 4). Moreover, we can see in the experimental results that some bands, which are far below the Fermi level, depend on the choice of the photon energy (\textit{k$_z$}  value). This is expected because although ${\mathrm{Nb}}_{3}{\mathrm{I}}_{8}$ is a 2D material, the experimental sample is technically a three-dimensional bulk material. Importantly, our calculations show that the flat and weakly dispersing bands arise from the 2D breathing kagome plane of the Nb atoms. \\

In summary, we carried out ARPES measurements on a semiconducting kagome material ${\mathrm{Nb}}_{3}{\mathrm{I}}_{8}$ that can be exfoliated down to 2D limit (see Supplementary Note 7 and Supplementary Fig. 10) and discover flat as well as weakly dispersing bands that arise from the breathing kagome lattice of the $\mathrm{Nb}$ atoms. Computational results using DFT support our observation. These bands have 2D nature as expected for a 2D material based on our photon-energy-dependent measurements and show a strong dependence on the polarization of the photon beam. Our study points to ${\mathrm{Nb}}_{3}{\mathrm{I}}_{8}$ being an important ground to study the interplay among geometry and electronic correlations in 2D materials. Given that the monolayer of this compound is predicted to be FM and the exfoliation to stable few-layers to monolayer limit already reported to be achieved, it could throw magnetism as well in the mix. \\

\noindent\textit{Note:} During the preparation of the manuscript, the authors became aware of similar work on a similar compound ${\mathrm{Nb}}_{3}{\mathrm{Cl}}_{8}$ \cite{Nb3Cl8}.\\

\noindent\textit{Acknowledgments - }
M. N. acknowledges the support from the National Science Foundation (NSF) CAREER award DMR-1847962, the Air Force Office of Scientific Research MURI Grant No. FA9550-20-1-0322, and NSF Partnerships for Research and Education in Materials (PREM) Grant DMR-2121953. X. X., T. C., J.-H. C., \& J. Y. acknowledge the support from UW Molecular Engineering Materials Center, a NSF Materials Research Science and Engineering Center (Grant No. DMR-1719797). H.V. was supported by NSF PREM Grant DMR-2121953.   This research used resources of the ALS, which is a DOE Office of Science User Facility under Contract No. DE-AC02-05CH11231.\\

\clearpage
\onecolumngrid
\setcounter{figure}{0}
\renewcommand{\figurename}{\textbf{Supplementary Fig.}}
\renewcommand{\thefigure}{{\textbf{\arabic{figure}}}}
\renewcommand{\tablename}{Supplementary Table}
\renewcommand{\thetable}{\arabic{table}}
\def\bibsection{\refname}
\renewcommand{\refname}{\noindent\textbf{Supplementary References}}
\begin{center}
\textbf{Supplemental Information}
\end{center}

\noindent\textbf{Supplementary Note 1. Crystal structure and sample characterization}\\
$\mathrm{Nb}_{3}\mathrm{I}_{8}$ has a layered crystal structure (rhombohedral space group no. 166) with intralayer covalent bonding and interlayer van der Waal bonding along  the crystallographic $\mathrm{c}$-axis. A side view of the bulk crystal structure is shown in Supplementary Fig. \ref{FigS1}a where a unit cell consists of sixtuple layers. Nb atoms are sandwiched in between two sheets of I atoms with I1, I3 atoms forming the top sheet and I2, I4 forming the bottom sheet. Top view of the crystal structure is presented in Supplementary Fig. \ref{FigS1}b. Nb atoms form a breathing kagome lattice [see Supplementary Fig. \ref{FigS1}c]. An schematic of breathing kagome lattice is shown in Supplementary Fig. \ref{FigS1}e, where the alternating triangles are of unequal size unlike in conventional kagome lattice in which all the triangles are of equal size [Supplementary Fig. \ref{FigS1}(d)].
\begin{figure*} [h!]
\includegraphics[width=0.88\textwidth]{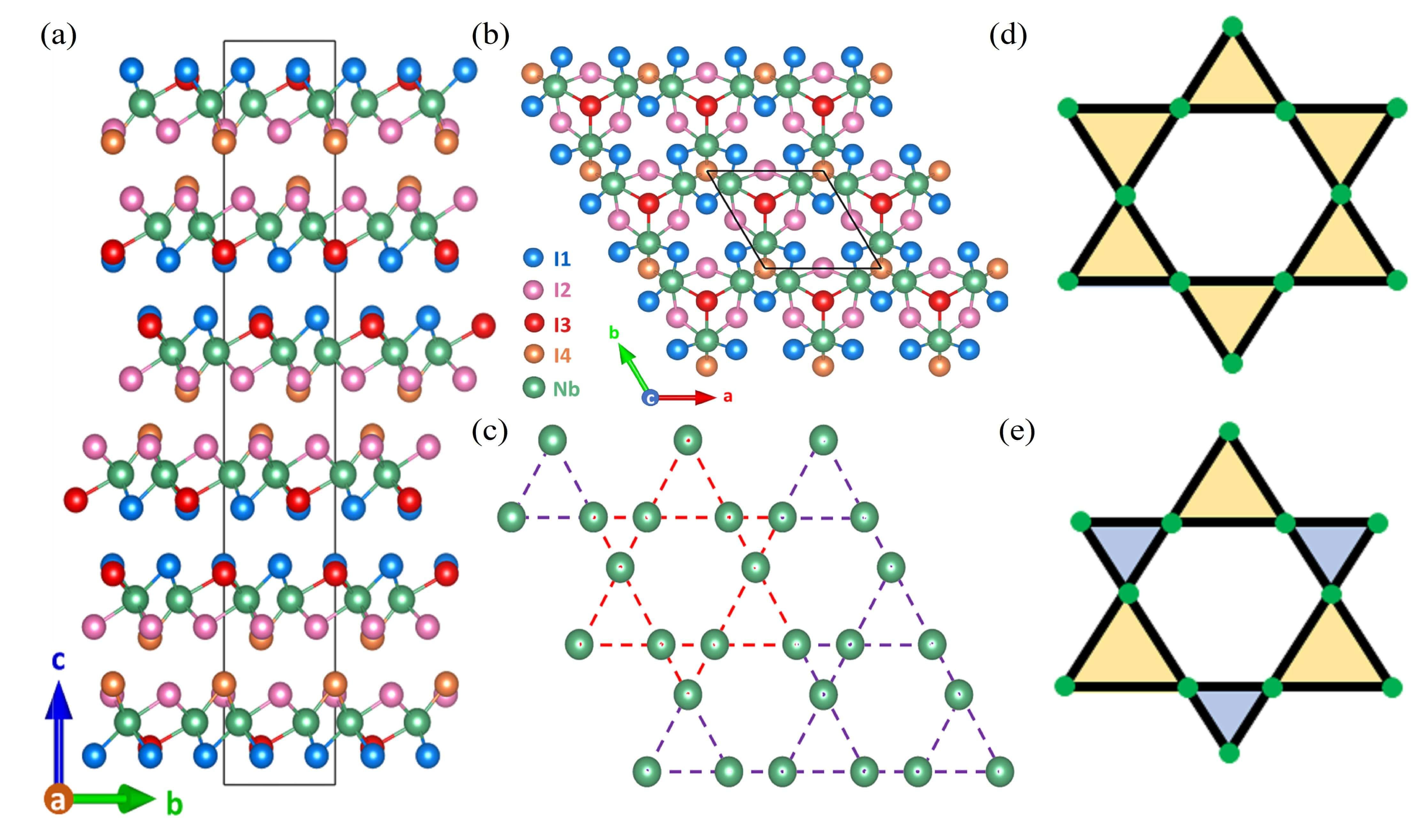}
\caption{\textbf{Crystal structure and kagome geometry in ${\mathrm{\textbf{Nb}}}_{3}{\mathrm{\textbf{I}}}_{8}$.} \textbf{a} Side view and \textbf{b} top view of the crystal structure. \textbf{c} Breathing kagome plane formed by Nb atoms. \textbf{d-e} Schematic of conventional kagome and breathing kagome lattices, respectively.}
\label{FigS1}
\end{figure*}

As mentioned in the method section of the main text, high-quality single crystals of ${\mathrm{Nb}}_{3}{\mathrm{I}}_{8}$ used in this study were grown by employing chemical vapor transport method with Iodine used as a transport agent.  0.2303 g High purity $\mathrm{Nb}$ powder (99.99\%, 325mesh from Alfa Aesar) and 0.8021 g $\mathrm{Iodine}$ (99.999\% from Aldrich) were sealed inside a quartz tube (12.75 mm OD $\times$ 10.5 mm  ID $\times$ 12 cm) under vacuum ($<$ 20 mtorr). The tube was loaded into a dual zone furnace and ramped for 20 hours to reach 652 $\degree$C on the precursor side and 650 $\degree$C on the growth side. After 70 hours growth, the tube and the furnace were allowed to naturally cool down to obtain flat and shiny crystals. The crystal structure was examined using X-ray diffraction (XRD). XRD on a ${\mathrm{Nb}}_{3}{\mathrm{I}}_{8}$ crystal grown in the same batch as the ARPES measurements is presented in Supplementary Fig. \ref{FigS2}a. Sharp $\mathrm{(000l)}$ peaks are observed indicating that the crystal surface is oriented in the $\mathrm{ab}$-plane. The compositional homogeneity of the crystals were verified using scanning electron microscopy (SEM) and energy dispersive X-ray spectroscopy analysis (EDS) performed on ${\mathrm{Nb}}_{3}{\mathrm{I}}_{8}$ flakes exfoliated on a $\mathrm{Si}\mathrm{O}_2/\mathrm{Si}$ substrate [see  Supplementary Fig. \ref{FigS2}b]. The quantitative elemental analysis shows a $\mathrm{Nb}:\mathrm{I}$ ratio of 8.24:22.31 = 2.96:8 [see table \ref{T1}].\\

\begin{figure*} [h!]
\includegraphics[width=0.80\textwidth]{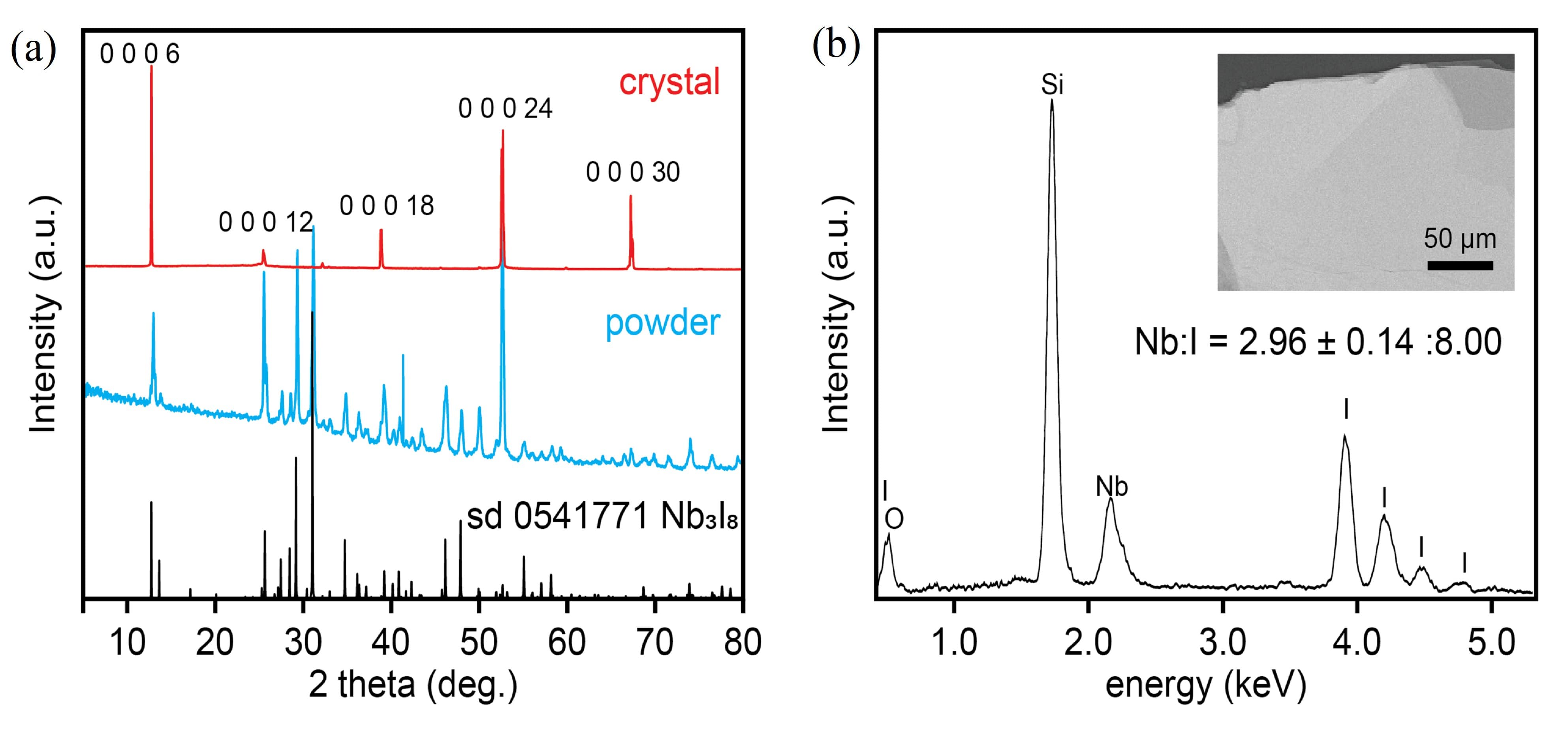}
\caption{\textbf{Sample characterization.} \textbf{a} X-ray Diffraction (XRD) on powdered (cyan) and single crystal (red) ${\mathrm{Nb}}_{3}{\mathrm{I}}_{8}$. \textbf{b} Energy-dispersive X-ray spectroscopy (EDS) of ${\mathrm{Nb}}_{3}{\mathrm{I}}_{8}$ flakes exfoliated on a $\mathrm{Si}\mathrm{O}_2/\mathrm{Si}$ substrate.}
\label{FigS2}
\end{figure*}

\begin{table}[h!]
\caption{EDS analysis of ${\mathrm{Nb}}_{3}{\mathrm{I}}_{8}$.}
\label{T1}
\begin{tabular}{|c|c|c|c|c|c|}
\hline
\textbf{Element} & \textbf{Atomic Number} & \textbf{Series} & \textbf{Wt\%} & \textbf{Wt\% Sigma} & \textbf{Atomic\%} \\
\hline
Silicon & 14 & K Series & 29.86 & 1.26 & 54.20 \\
\hline
Iodine & 53 & L Series & 55.53 & 1.64 & 22.31 \\
\hline
Oxygen & 8 & K Series & 4.78 & 0.8 & 15.24 \\
\hline
Niobium & 41 & L Series & 15.02 & 0.56 & 8.24\\
\hline
\end{tabular}
\end{table}

\noindent\textbf{Supplementary Note 2. First-principle calculations}\\
First-principles calculations were performed, for the bulk stable state (Antiferromagnetic, interlayer spin configuration with alternate layers having opposite spins) and monolayer stable state (FM), utilizing the density functional theory approach \cite{SIDFT1, SIDFT2} as implemented in the {\sc Quantum ESPRESSO} package \cite{SIQE} by using norm-conserving scalar relativistic ONCVPSP (Optimized Norm-Conserving Vanderbilt Pseudopotential) \cite{SIONCVP1, SIONCVP2} and Perdew-Burke-Ernzerhof-type exchange-correlational functional \cite{SIPBE} and with van der Waals correction included in the D2 formalism \cite{SID2}. A conventional unit cell with six layers with $a = \mathrm{7.600~\AA}$ and $c = \mathrm{41.715~\AA}$ \cite{SIMagonov} was used for the bulk calculations, and a single ${\mathrm{Nb}}_{3}{\mathrm{I}}_{8}$ layer with $c = \mathrm{52.500~\AA}$ was considered for monolayer calculations. Interlayer AFM was considered with each consecutive layers in six layers having alternate spins.  Hubbard potential $U = \mathrm{2~eV}$ on $\mathrm{Nb}~d$ orbitals was used to account for the on-site interaction \cite{SIJiangNb3X8}. Convergence tests were run to justify a kinetic energy cutoff, $ecutwfc = \mathrm{85~Ry}$, and an 8 x 8 x 1 Monkhorst-Pack k-point grid. We used a convergence threshold (for self-consistency) of $\mathrm{1\times10^{-8}~Ry}$ .\\

In Supplementary Fig. \ref{FigS3}, DFT calculations for bulk as well as monolayer ${\mathrm{Nb}}_{3}{\mathrm{I}}_{8}$ are presented. For comparison with experimental data, we set the mid-band gap to be the zero-energy level. The overall band structures of bulk and monolayer look similar. Direct Kohn-Sham band gap at the $\mathrm{K}$ point for bulk is $\sim \mathrm{520~meV}$ and that of monolayer is $\sim \mathrm{490~meV}$ indicating semiconducting nature of ${\mathrm{Nb}}_{3}{\mathrm{I}}_{8}$ in both three- and two-dimensions. We calculated the mirror operator eigenvalues of -1, 1, 1, 1, 1, -1, 1, and -1 for the bands c3, c2, c1, v1, v2, v3, v4, and v5 bands (see Supplementary Fig. \ref{FigS4} for monolayer Kohn-Sham bands), respectively. With the inclusion of spin-orbit coupling (SOC), the band degeneracy of these bands at the $\mathrm{\Gamma}$ point is broken, indicative of potential topological character. For c1 and v1, no band inversion occurs, therefore these bands are non-topological. For bands c2, c3, v2, v3, v4, and v5, since any of these SOC-gapped bands cannot be completely isolated in energy by shifting the Fermi level (unlike band v1), quantized non-zero Chern numbers do not exist in this band structure by rigidly shifting the Fermi level alone. \\

\begin{figure*} [h!]
\includegraphics[width=0.85\textwidth]{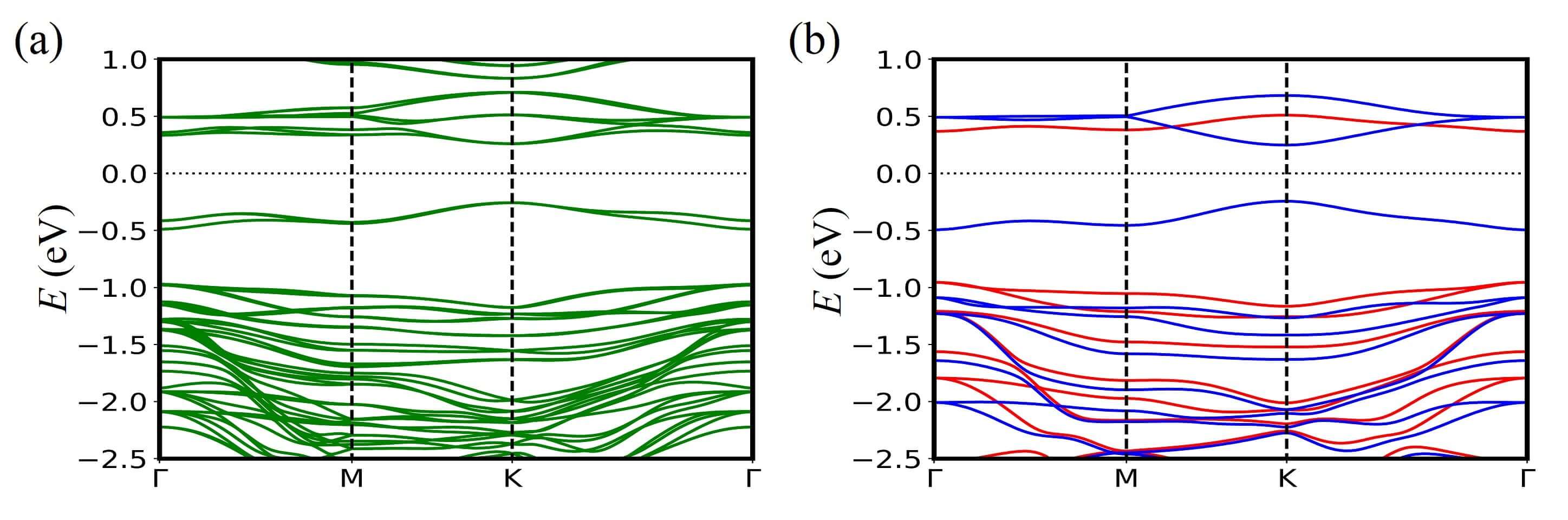}
\caption{\textbf{Bulk and monolayer band calculations.} \textbf{a} DFT band structure calculation for  bulk ${\mathrm{Nb}}_{3}{\mathrm{I}}_{8}$. For bulk, majority- and minority-spin bands are degenerate. \textbf{b} DFT band structure calculation for monolayer ${\mathrm{Nb}}_{3}{\mathrm{I}}_{8}$.  Blue and red denote majority- and minority-spin bands, respectively.}
\label{FigS3}
\end{figure*}

\begin{figure*} [h!]
\includegraphics[width=0.85\textwidth]{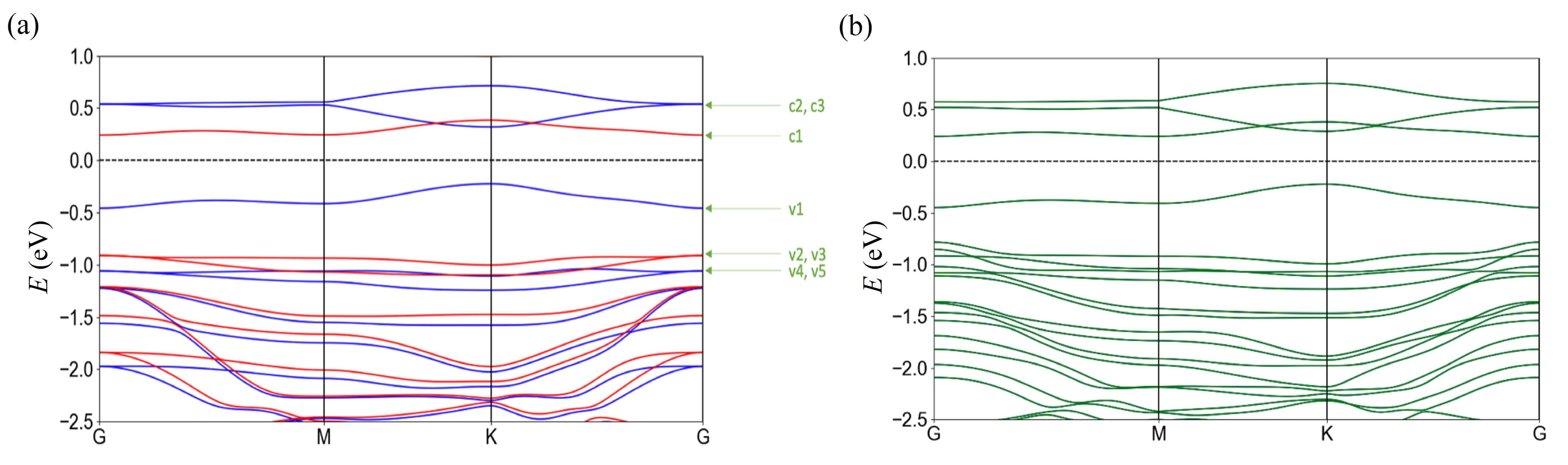}
\caption{\textbf{Kohn-Sham bands of monolayer $\mathrm{\textbf{Nb}}_3\mathrm{\textbf{I}}_8$.} \textbf{a} Calculated Kohn-Sham bands without considering SOC. \textbf{b} Kohn-Sham bands with SOC included.}
\label{FigS4}
\end{figure*}

In Supplementary Fig. \ref{FigS5}, we present the calculations for bulk ${\mathrm{Nb}}_{3}{\mathrm{I}}_{8}$ at different $k_z$ planes (0 and $\pi$). Note that these calculations have been implemented in {\sc VASP} \cite{VASP}. The overall band structure remains similar to the one obtained by implementing in {\sc Quantum ESPRESSO}. Also, the band structure within the bands of interest (flat and weakly dispersing bands that lie well within $\mathrm{1.5~eV}$) does not have significant difference for $k_z = 0$ and $k_z = \pi$ planes.\\

\begin{figure*} [h!]
\includegraphics[width=0.85\textwidth]{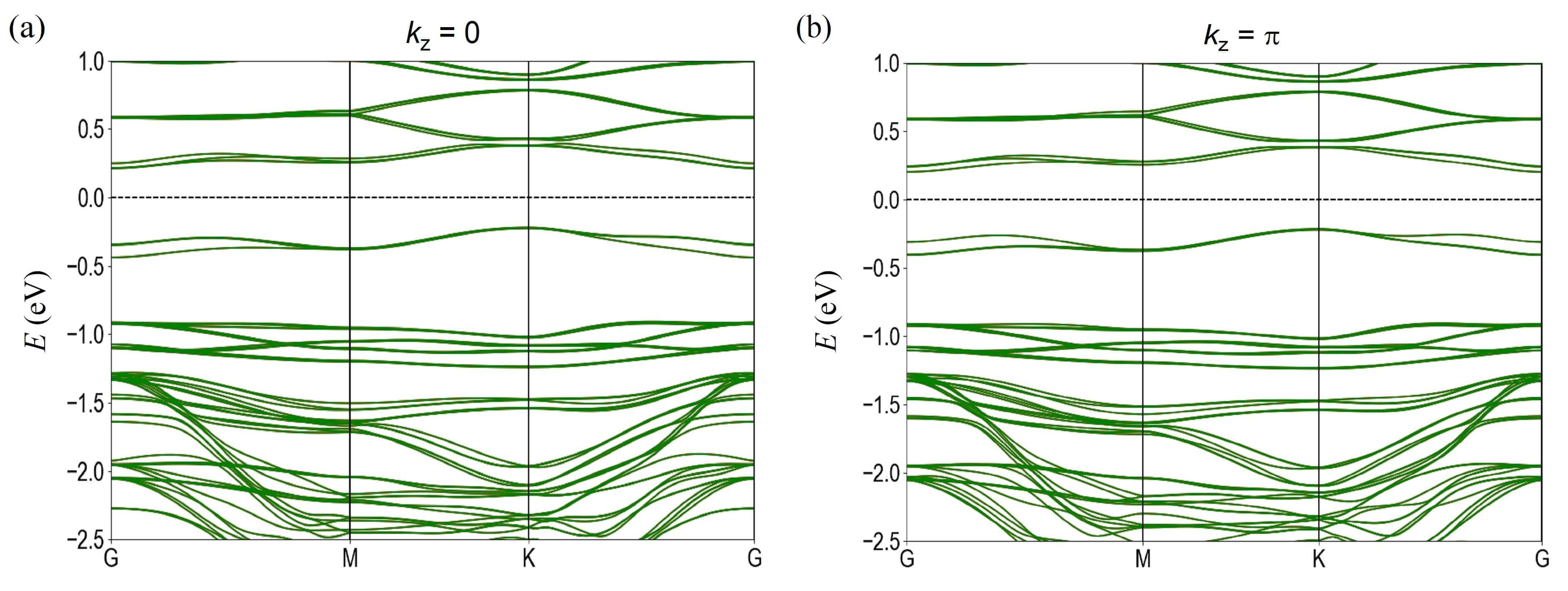}
\caption{\textbf{Bulk bands calculated for different $k_z$ planes.} DFT band calculations for  bulk ${\mathrm{Nb}}_{3}{\mathrm{I}}_{8}$ for \textbf{a} $k_z = 0$ and \textbf{b} $k_z = \pi$ planes. The calculations are performed using {\sc VASP} software \cite{SIKresse1996}.}
\label{FigS5}
\end{figure*}

\noindent\textbf{Supplementary Note 3. Experimental details}\\
Experimental measurements using the angle-resolved photoemission spectroscopy (ARPES) technique were carried out at the Advanced Light Source Beamline 4.0.3 equipped with Scienta R8000 hemispherical electron analyzer in the Berkeley National Laboratory. Single crystals of ${\mathrm{Nb}}_{3}{\mathrm{I}}_{8}$ were cleaved in-situ in ultra-high vacuum conditions to achieve clean and fresh surface. Measurements were done at a temperature of 260 K in order to dodge the charging effects. For the whole measurement period, the sample surface was stable and showed no sign of degradation. Energy and angular resolutions were set to be better than 20 meV and 0.2$\degree$, respectively. Photon beam with monoenergy in the range 30-124 eV was used with linear horizontal (LH) and linear vertical polarization (LV). \\
\begin{figure*} [h!]
\includegraphics[width=0.5\textwidth]{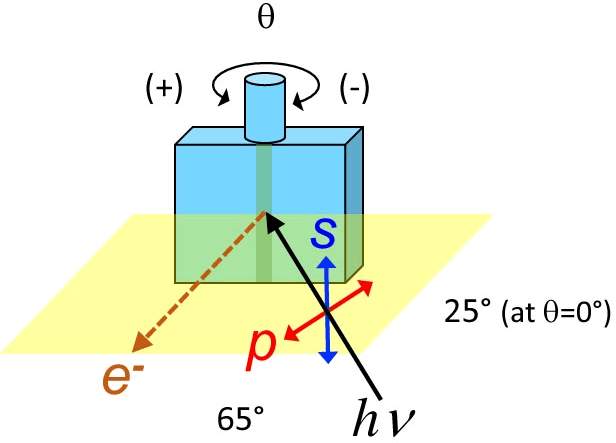}
\caption{Experimental geometry used in the ARPES measurements.}
\label{FigS6}
\end{figure*}
An experimental geometry for the ARPES measurements is shown in Supplementary Fig. \ref{FigS6}. At normal emission, the beam is incident at a 25$\degree$ grazing angle to the sample, i.e. the angle between the x-rays and Scienta lens axis is fixed at 65$\degree$. The Scienta electron analyzer is mounted with a vertical slit and a polar map rotates the sample about a vertical axis. LH (LV) polarization oscillates the electrons in the undulator in the horizontal (vertical) plane and so the x-ray polarization vector is in the horizontal  (vertical) plane of the incident x-ray propagation direction. When the sample tilt angle is zero, then LV polarization gives purely s-polarization (polarization vector in the plane of the sample surface). LH polarization has both s- and p-polarization components, which depend on the polar angle ($\theta$) as: $e_p = cos(\theta+25\degree)$ and $e_s = sin(\theta+25\degree)$. For our polarization dependent cuts, the value of $\theta$ is -4$\degree$.\\

\noindent\textbf{Supplementary Note 4. Photon energy dependent dispersion maps}\\
In order to show the two-dimensional (2D) nature of the flat bands, we show the photon energy dependent dispersion maps along the $\mathrm{\overline{M}} - \mathrm{\overline{\Gamma}} - \mathrm{\overline{M}}$ direction in Supplementary Fig. \ref{FigS7}. Similar to what was observed with the photon energy variation in the main text, the flat bands do not change their dispersion with changing photon energy (only the photoemission intensity is changing). This confirms that these bands are indeed of 2D nature. On the other hand, the dispersing bands seem to be dependent on photon energy showing their three-dimensional bulk origination.\\
\begin{figure*} [h!]
\includegraphics[width=1\textwidth]{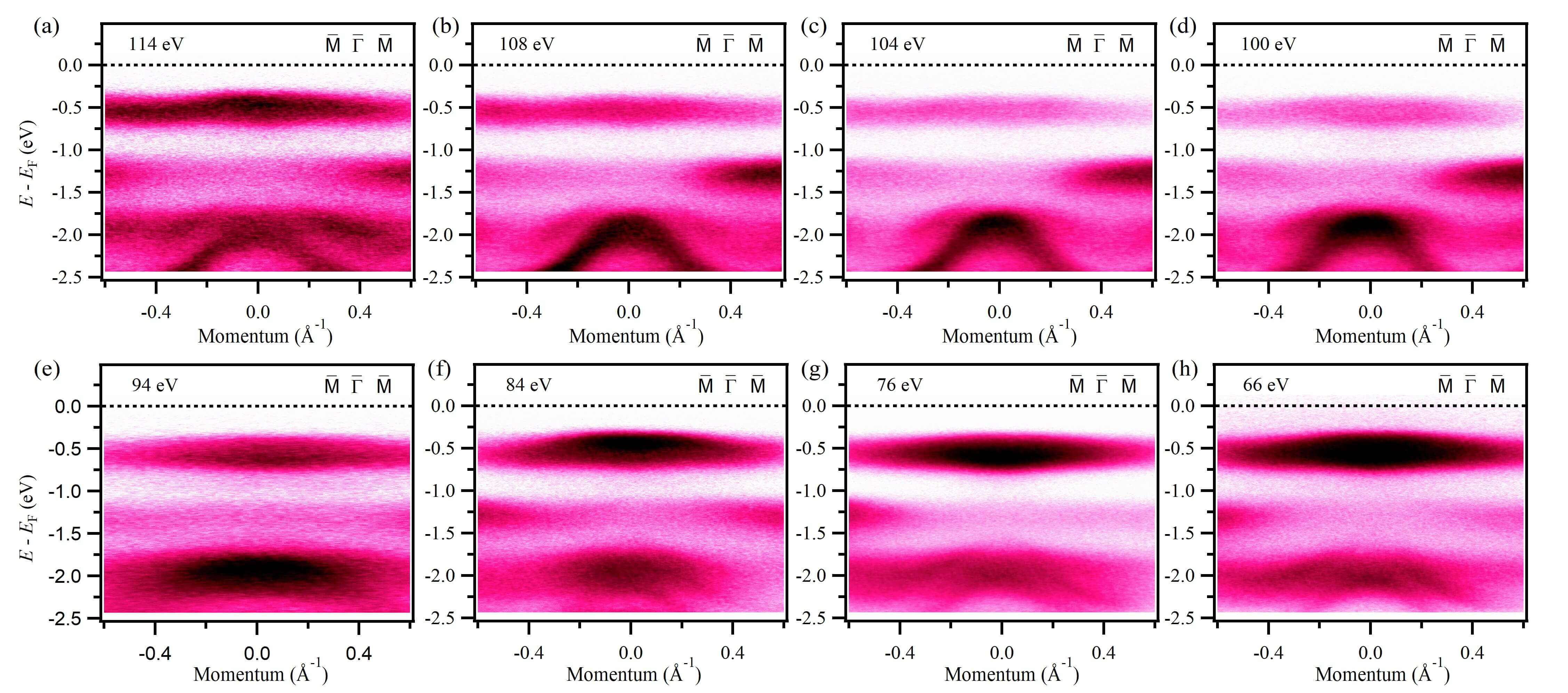}
\caption{\textbf{Photon energy-dependent dispersion maps.} \textbf{a-h} Dispersion maps along the $\mathrm{\overline{M}} - \mathrm{\overline{\Gamma}} - \mathrm{\overline{M}}$ direction for different value of photon mono-energy. The energy values are noted on each plots.}
\label{FigS7}
\end{figure*}

\noindent\textbf{Supplementary Note 5. Orbital resolved band structure calculations}\\
\begin{figure*} [h!]
\includegraphics[width=0.90\textwidth]{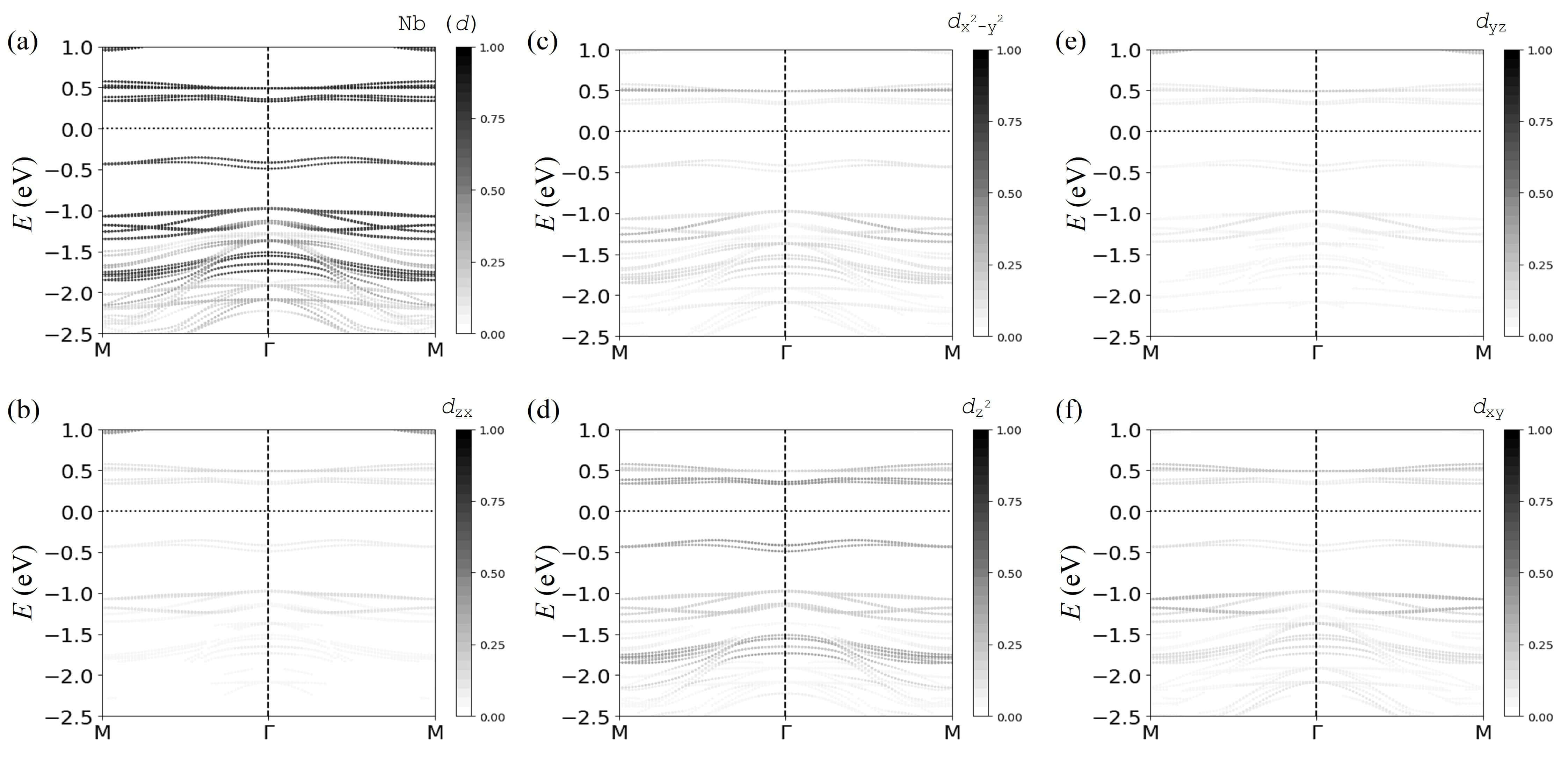}
\caption{\textbf{Contribution from individual $\mathrm{\textbf{Nb}}~d$ orbitals.} \textbf{a} Dispersion map along the $\mathrm{M} - \mathrm{\Gamma} - \mathrm{M}$ direction with total contribution from Nb \textit{d} orbitals. \textbf{b-f} Band structure showing individual contributions from $d_{zx}$, $d_{x^2-y^2}$, $d_{z^2}$, $d_{yz}$, and $d_{xy}$ orbitals, respectively. }
\label{FigS8}
\end{figure*}
Supplementary Fig. \ref{FigS8}a shows the total contribution from Noibium \textit{d} in the electronic structure along $\mathrm{M} - \mathrm{\Gamma} - \mathrm{M}$. We can see strong contribution of Nb \textit{d} to the bands A - C. Subsequent plots in Supplementary Figs. \ref{FigS8}b-f present the individual contribution from $d_{zx}$, $d_{x^2-y^2}$, $d_{z^2}$, $d_{yz}$, and $d_{xy}$ orbitals, respectively. It can be seen that the bands A and B have strong dominance from $d_{z^2}$ orbitals. The bandset C, on the other hand, seem to have significant contribution coming from all orbitals with majority contribution coming from  $d_{xy}$ orbitals.\\

\noindent\textbf{Supplementary Note 6. Energy distribution curves (EDCs)}\\
In Supplementary Fig. \ref{FigS9}, we present the EDCs (both raw and fits using voigt function) for the plots used in main text Figs. 3a and 4d. We can observe that the intensity of the flat and weakly dispersing bands coming from Nb breathing kagome plane depend upon the polarization of the incident photon beam. The bands around 500 meV (A and B), which have dominant $d_{z^2}$ orbital character appear strong when the incident beam is LH polarized. On the other hand, the bands below 1.2 eV (bandset C) with major $d_{xy}$ orbital contribution are enhanced when the incident photon beam is LV polarization. \\
\begin{figure*} [h!]
\includegraphics[width=0.90\textwidth]{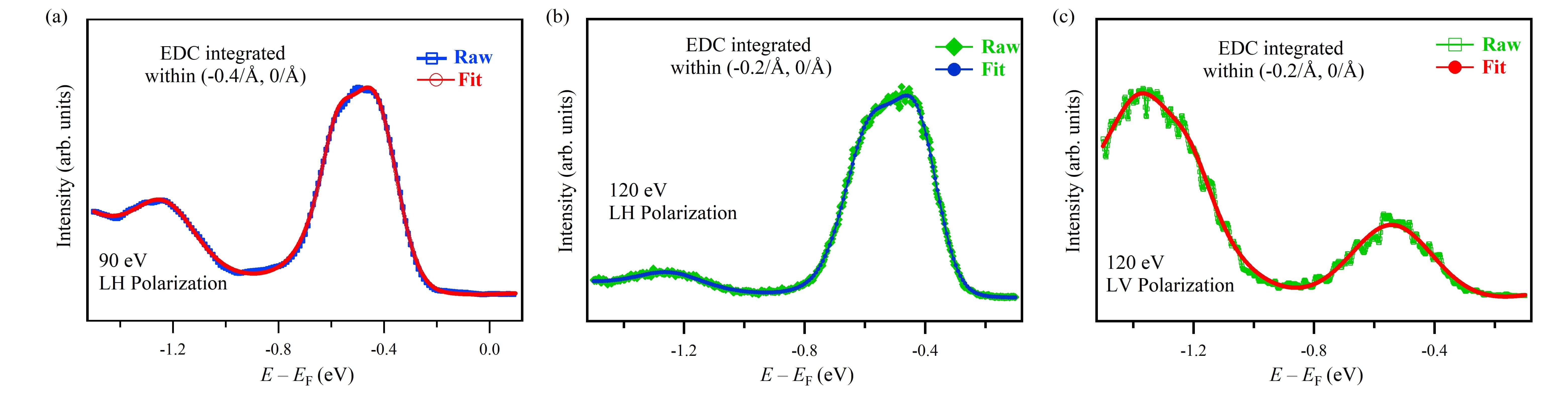}
\caption{\textbf{Energy distribution curves and Voigt fits.} EDCs and respective fits using voigt function for \textbf{a} 90 eV linear horizontal polarized light, \textbf{b} 120 eV linear horizontal polarized light, and \textbf{c} 120 eV linear vertical polarized light.}
\label{FigS9}
\end{figure*}

\noindent\textbf{Supplementary Note 7. Optical and atomic force microscopy images}\\
In Supplementary Fig. \ref{FigS10}, we present the images of the exfoliated ${\mathrm{Nb}}_{3}{\mathrm{I}}_{8}$ samples. Panel (a) shows the optical image of the crystal that is exfoliated and the panels (b) and (c) show the optical image and atomic force microscopy images after exfoliation onto $\mathrm{SiO}_2/\mathrm{Si}$ substrate. Our images reveal that the bulk ${\mathrm{Nb}}_{3}{\mathrm{I}}_{8}$ crystals can be exfoliated into thinner flakes, thereby indicating the potential utilization in two-dimensional heterostructure studies and applications.\\
\begin{figure*} [h!]
\includegraphics[width=0.85\textwidth]{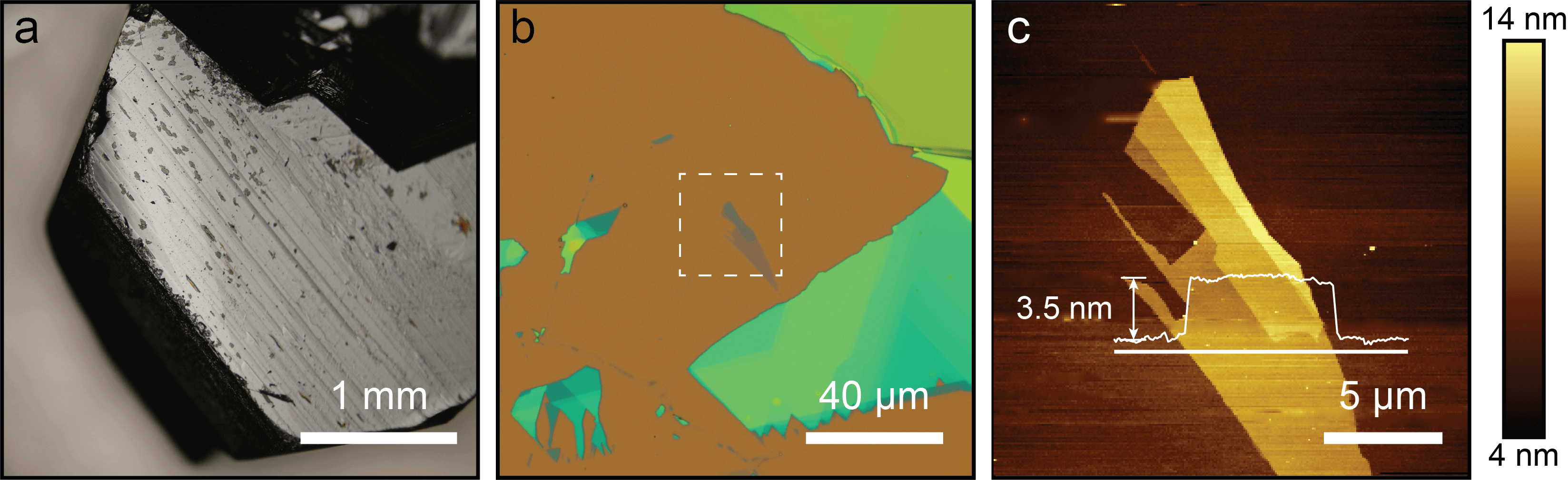}
\caption{\textbf{Exfoliation of ${\mathrm{\textbf{Nb}}}_{3}{\mathrm{\textbf{I}}}_{8}$.} \textbf{a} Optical image of crystal before exfoliation. \textbf{b,c} Optical image and atomic force microscopy image of exfoliated ${\mathrm{Nb}}_{3}{\mathrm{I}}_{8}$ samples.}
\label{FigS10}
\end{figure*}


\begin{thebibliography}{50}
\bibitem{Kida} Kida, T. et al. The giant anomalous Hall effect in the ferromagnet $\mathrm{Fe}_3\mathrm{Sn}_2$—a frustrated kagome metal. \href{https://doi.org/10.1088/0953-8984/23/11/112205}{\textit{J. Phys. Condens. Mater.} \textbf{23,} 112205 (2011)}.

\bibitem{Han} Han, T.-H. et al. Fractionalized excitations in the spin-liquid state of a kagome-lattice antiferromagnet.  \href{https://doi.org/10.1038/nature11659}{\textit{Nature} \textbf{492,} 406 (2012)}.

\bibitem{Nakatsuji} Nakatsuji, S., Kiyohara, N. \& Higo, T. Large anomalous Hall effect in a non-collinear antiferromagnet at room temperature.  \href{https://doi.org/10.1038/nature15723}{\textit{Nature} \textbf{527,} 212 (2015)}.


\bibitem{Fe3Sn2_Lin} Lin, Z. et al. Flatbands and Emergent Ferromagnetic Ordering in ${\mathrm{Fe}}_{3}{\mathrm{Sn}}_{2}$ Kagome Lattices. \href{https://doi.org/10.1103/PhysRevLett.121.096401}{\textit{Phys. Rev. Lett.} \textbf{121,} 096401 (2018)}.

\bibitem{Yin1} Yin, J.-X. et al. Giant and anisotropic many-body spin–orbit tunability in a strongly correlated kagome magnet. \href{https://doi.org/10.1038/s41586-018-0502-7}{\textit{Nature} \textbf{562,} 91 (2018)}.

\bibitem{Yin2} Yin, J.-X. et al. Negative flat band magnetism in a spin–orbit-coupled correlated kagome magnet. \href{https://doi.org/10.1038/s41567-019-0426-7} {\textit{Nat. Phys.} \textbf{15,} 443 (2019)}.


\bibitem{Yang} Yang, S.-Y. et al. Giant, unconventional anomalous Hall effect in the metallic frustrated magnet candidate, $\mathrm{K}\mathrm{V}_3\mathrm{Sb}_5$.  \href{https://doi.org/10.1126/sciadv.abb6003}{\textit{Sci. Adv.} \textbf{6,} eabb6003 (2020)}.

\bibitem{Tb166} Yin, J.-X. et al. Quantum-limit Chern topological magnetism in $\mathrm{Tb}\mathrm{Mn}_6\mathrm{Sn}_6$. \href{https://doi.org/10.1038/s41586-020-2482-7}{\textit{Nature} \textbf{583,} 533 (2020)}.

\bibitem{Lin} Lin, Z. et al. Dirac fermions in antiferromagnetic  $\mathrm{FeSn}$ kagome lattices with combined space inversion and time-reversal symmetry. \href{https://doi.org/10.1103/PhysRevB.102.155103}{\textit{Phys. Rev. B} \textbf{102,} 155103 (2020)}.

\bibitem{Ghimire} Ghimire, N. J. et al. Competing magnetic phases and fluctuation-driven scalar spin chirality in the kagome metal $\mathrm{Y}\mathrm{Mn}_6\mathrm{Sn}_6$.  \href{https://doi.org/10.1126/sciadv.abe2680}{\textit{Sci. Adv.} \textbf{6,} eabe2680 (2020)}.





\bibitem{Dirac} Mazin, I. I. et al. Theoretical prediction of a strongly correlated Dirac metal. \href{https://doi.org/10.1038/ncomms5261}{\textit{Nat. Commun.} \textbf{5,} 4261 (2014)}.

\bibitem{CoSn_1} Liu, Z. et al. Orbital-selective Dirac fermions and extremely flat bands in frustrated kagome-lattice metal $\mathrm{CoSn}$. \href{https://doi.org/10.1038/s41467-020-17462-4}{\textit{Nat. Commun.} \textbf{11,} 4002 (2020)}.

\bibitem{CoSn_2} Kang, M. et al. Topological flat bands in frustrated kagome lattice $\mathrm{CoSn}$.  \href{https://doi.org/10.1038/s41467-020-17465-1}{\textit{Nat. Commun.} \textbf{11,} 4004 (2020)}.

\bibitem{FeSn} Kang, M. et al. Dirac fermions and flat bands in the ideal kagome metal $\mathrm{FeSn}$.  \href{https://doi.org/10.1038/s41563-019-0531-0}{\textit{Nat. Mater.} \textbf{19,} 163 (2020)}.

\bibitem{Y166} Li, M. et al. Dirac cone, flat band and saddle point in kagome magnet $\mathrm{Y}\mathrm{Mn}_6\mathrm{Sn}_6$.  \href{https://doi.org/10.1038/s41467-021-23536-8}{\textit{Nat. Commun.} \textbf{12,} 3129 (2021)}.

\bibitem{Tang} Tang, E., Mei, J.-W. \& Wen, X.-G. High-Temperature Fractional Quantum Hall States.  \href{https://doi.org/10.1103/PhysRevLett.106.236802}{\textit{Phys. Rev. Lett.} \textbf{106,} 236802 (2011)}.

\bibitem{Xu} Xu, G., Lian, B. \& Zhang, S.-C. Intrinsic Quantum Anomalous Hall Effect in the Kagome Lattice ${\mathrm{Cs}}_{2}{\mathrm{LiMn}}_{3}{\mathrm{F}}_{12}$. \href{https://doi.org/10.1103/PhysRevLett.115.186802}{\textit{Phys. Rev. Lett.} \textbf{115,} 186802 (2015)}.

\bibitem{Fe3Sn2_2} Ye, L. et al. Massive Dirac fermions in a ferromagnetic kagome metal. \href{https://doi.org/10.1038/nature25987}{\textit{Nature} \textbf{555,} 638 (2018)}.

\bibitem{Bolens} Bolens, A. \& Nagaosa, N. Topological states on the breathing kagome lattice. \href{https://doi.org/10.1103/PhysRevB.99.165141}{\textit{Phys. Rev. B} \textbf{99,} 165141 (2019)}.

\bibitem{Ezawa} Ezawa, M. Higher-Order Topological Insulators and Semimetals on the Breathing Kagome and Pyrochlore Lattices. \href{https://doi.org/10.1103/PhysRevLett.120.026801}{\textit{Phys. Rev. Lett.} \textbf{120,} 026801 (2018)}.

\bibitem{YLi} Li, Y., Liu, C., Zhao, G.-D., Hu, T. \& Ren, W. Two-dimensional multiferroics in a breathing kagome lattice. \href{https://doi.org/10.1103/PhysRevB.104.L060405}{\textit{Phys. Rev. B} \textbf{104,} L060405 (2021)}.

\bibitem{ARPES1} Damascelli, A., Hussain, Z. \& Shen, Z.-X. Angle-resolved photoemission studies of the cuprate superconductors.  \href{https://doi.org/10.1103/RevModPhys.75.473}{\textit{Rev. Mod. Phys.} \textbf{75,} 473 (2003)}.

\bibitem{ARPES2} Damascelli, A. Probing the Electronic Structure of Complex Systems by ARPES.  \href{https://doi.org/10.1238/Physica.Topical.109a00061}{\textit{Phys. Scr.} \textbf{T109,} 61 (2004)}.

\bibitem{ARPES3} Lv, B., Qian, T. \& Ding, H. Angle-resolved photoemission spectroscopy and its application to topological materials. \href{https://doi.org/10.1038/s42254-019-0088-5}{\textit{Nat. Rev. Phys.} \textbf{1,} 609 (2019)}.

{\bibitem{Tanaka} Tanaka, H. et al. Three-dimensional electronic structure in ferromagnetic ${\mathrm{Fe}}_{3}{\mathrm{Sn}}_{2}$ with breathing kagome bilayers. \href{https://doi.org/10.1103/PhysRevB.101.161114}{\textit{Physical Review B} \textbf{101,} 161114 (2020)}.}

\bibitem{Er166} Dhakal, G. et al. Anisotropically large anomalous and topological Hall effect in a kagome magnet.  \href{https://doi.org/10.1103/PhysRevB.104.L161115}{\textit{Phys. Rev. B} \textbf{104,} L161115 (2021)}.

\bibitem{Ho166}  Kabir, F. et al. Unusual magnetic and transport properties in $\mathrm{Ho}\mathrm{Mn}_6\mathrm{Sn}_6$ kagome magnet. \href{https://doi.org/10.1103/PhysRevMaterials.6.064404}{\textit{Phys. Rev. Materials} \textbf{6,} 064404 (2022)}.

\bibitem{RV166} Peng, S. et al. Realizing Kagome Band Structure in Two-Dimensional Kagome Surface States of $R{\mathrm{V}}_{6}{\mathrm{Sn}}_{6}$ ($R=\mathrm{Gd}$, Ho). \href{https://doi.org/10.1103/PhysRevLett.127.266401}{\textit{Phys. Rev. Lett.} \textbf{127,} 266401 (2021)}.

\bibitem{Ortiz} Ortiz, B. R. et al. $\mathrm{Cs}{\mathrm{V}}_{3}{\mathrm{Sb}}_{5}$: A ${\mathbb{Z}}_{2}$ Topological Kagome Metal with a Superconducting Ground State. \href{https://doi.org/10.1103/PhysRevLett.125.247002}{\textit{Phys. Rev. Lett.} \textbf{125,} 247002 (2020)}.

\bibitem{Magonov}  Magonov, S. N. et al. Scanning tunneling and atomic force microscopy study of layered transition metal halides ${\mathrm{Nb}}_{3}{\mathrm{X}}_{8}$ ($\mathrm{X}$ = $\mathrm{Cl}$, $\mathrm{Br}$, $\mathrm{I}$).  \href{https://doi.org/10.1021/ja00059a053}{\textit{J. Am. Chem. Soc.} \textbf{115,} 2495 (1993)}.

\bibitem{JiangNb3X8} Jiang, J. et al. Exploration of new ferromagnetic, semiconducting and biocompatible ${\mathrm{Nb}}_{3}{\mathrm{X}}_{8}$ ($\mathrm{X}$ = $\mathrm{Cl}$, $\mathrm{Br}$ or $\mathrm{I}$) monolayers with considerable visible and infrared light absorption. \href{https://doi.org/10.1039/C6NR07231C}{\textit{Nanoscale} \textbf{9,} 2992 (2017)}.

\bibitem{Kim} Kim, B. J. et al. Structural and Electrical Properties of ${\mathrm{Nb}}_{3}{\mathrm{I}}_{8}$ Layered Crystal. \href{ https://doi.org/10.1002/pssr.201800448}{\textit{Phys. Status Solidi RRL} \textbf{13,} 1800448 (2019)}.

\bibitem{Oh} Oh, S. et al. Large-area synthesis of van der Waals two-dimensional material ${\mathrm{Nb}}_{3}{\mathrm{I}}_{8}$ and its infrared detection applications.  \href{https://doi.org/10.1016/j.jallcom.2020.154877}{\textit{J.  Alloys Compd.} \textbf{831,} 154877 (2020)}.

\bibitem{Yoon} Yoon, J. et al. Anomalous thickness-dependent electrical conductivity in van der Waals layered transition metal halide, ${\mathrm{Nb}}_{3}{\mathrm{Cl}}_{8}$. \href{https://doi.org/10.1088/1361-648X/ab832b}{\textit{J. Phys. Condens. Mater.} \textbf{32,} 304004 (2020)}.

\bibitem{Conte} Conte, F., Ninno, D. \& Cantele, G. Layer-dependent electronic and magnetic properties of ${\mathrm{Nb}}_{3}{\mathrm{I}}_{8}$.  \href{https://doi.org/10.1103/PhysRevResearch.2.033001}{\textit{Phys. Rev. Res.} \textbf{2,} 033001 (2020)}.

\bibitem{Conte2} Cantele, G., Conte, F., Zullo, L. \& Ninno, D. Tunable electronic and magnetic properties of thin ${\mathrm{Nb}}_{3}{\mathrm{I}}_{8}$ nanofilms: Interplay between strain and thickness. \href{https://doi.org/10.1103/PhysRevB.106.085418}{\textit{Phys. Rev. B} \textbf{106,} 085418 (2022)}.

\bibitem{RuiPeng} Peng, R. et al. Intrinsic anomalous valley Hall effect in single-layer $\mathrm{N}{\mathrm{b}}_{3}{\mathrm{I}}_{8}$. \href{https://doi.org/10.1103/PhysRevB.102.035412}{\textit{Phys. Rev. B} \textbf{102,} 035412 (2020)}.

\bibitem{DFT1}  Hohenberg P. \&  Kohn W.  Inhomogeneous Electron Gas.  \href{https://doi.org/10.1103/PhysRev.136.B864}{Phys. Rev. \textbf{136,} B864 (1964)}.

\bibitem{DFT2}  Kohn W. \&  Sham L. J. Self-Consistent Equations Including Exchange and Correlation Effects. \href{https://doi.org/10.1103/PhysRev.140.A1133}{Phys. Rev. \textbf{140,} A1133 (1965)}.

\bibitem{QE} Giannozzi, P. et al. QUANTUM ESPRESSO: a modular and open-source software project for quantum simulations of materials. \href{https://doi.org/10.1088/0953-8984/21/39/395502}{\textit{J. Phys. Condens. Mater.} \textbf{21,} 395502 (2009)}.

\bibitem{Kresse1996} Kresse, G. \& Furthmüller, J. Efficient iterative schemes for ab initio total-energy calculations using a plane-wave basis set. \href{https://doi.org/10.1103/PhysRevB.54.11169}{\textit{Phys. Rev. B} \textbf{54,} 11169 (1996)}.

\bibitem{ONCVP1} Hamann, D. R. Optimized norm-conserving Vanderbilt pseudopotentials. \href{https://doi.org/10.1103/PhysRevB.88.085117}{\textit{Phys. Rev. B} \textbf{88,} 085117 (2013)}.

\bibitem{ONCVP2} Hamann, D. R. Erratum: Optimized norm-conserving Vanderbilt pseudopotentials [Phys. Rev. B 88, 085117 (2013)]. \href{https://doi.org/10.1103/PhysRevB.95.239906}{\textit{Phys. Rev. B} \textbf{95,} 239906 (2017)}.

\bibitem{PBE} Perdew J. P., Burke K., \&  Ernzerhof M. Generalized Gradient Approximation Made Simple. \href{https://doi.org/10.1103/PhysRevLett.77.3865}{Phys. Rev. Lett. \textbf{77}, 3865 (1996)}.

\bibitem{D2} Grimme, S. Semiempirical GGA-type density functional constructed with a long-range dispersion correction. \href{https://doi.org/10.1002/jcc.20495}{\textit{J. Comput. Chem.} \textbf{27,} 1787 (2006)}.

\bibitem{Nb3Cl8} Sun, Z. et al. Observation of Topological Flat Bands in the Kagome Semiconductor $\mathrm{Nb}_3\mathrm{Cl}_8$. \href{https://doi.org/10.1021/acs.nanolett.2c00778}{\textit{Nano Lett.} \textbf{22,} 4596 (2022)}.
\end{thebibliography}

\begin{thebibliography}{50}
\bibitem{SIDFT1}  Hohenberg, P. \&  Kohn, W.  Inhomogeneous Electron Gas.  \href{https://doi.org/10.1103/PhysRev.136.B864}{\textit{Phys. Rev.} \textbf{136,} B864 (1964)}.

\bibitem{SIDFT2}  Kohn, W. \&  Sham, L. J. Self-Consistent Equations Including Exchange and Correlation Effects. \href{https://doi.org/10.1103/PhysRev.140.A1133}{\textit{Phys. Rev.} \textbf{140,} A1133 (1965)}.

\bibitem{SIQE} Giannozzi, P. et al. QUANTUM ESPRESSO: a modular and open-source software project for quantum simulations of materials. \href{https://doi.org/10.1088/0953-8984/21/39/395502}{\textit{J. Phys. Condens. Mater.} \textbf{21,} 395502 (2009)}.

\bibitem{SIONCVP1} Hamann, D. R. Optimized norm-conserving Vanderbilt pseudopotentials. \href{https://doi.org/10.1103/PhysRevB.88.085117}{\textit{Phys. Rev. B} \textbf{88,} 085117 (2013)}.

\bibitem{SIONCVP2} Hamann, D. R. Erratum: Optimized norm-conserving Vanderbilt pseudopotentials [Phys. Rev. B 88, 085117 (2013)]. \href{https://doi.org/10.1103/PhysRevB.95.239906}{\textit{Phys. Rev. B} \textbf{95,} 239906 (2017)}.

\bibitem{SIPBE} J. P. Perdew, K. Burke, \& M. Ernzerhof. Generalized Gradient Approximation Made Simple, \href{https://doi.org/10.1103/PhysRevLett.77.3865}{Phys. Rev. Lett. \textbf{77}, 3865 (1996)}.

\bibitem{SID2} Grimme, S. Semiempirical GGA-type density functional constructed with a long-range dispersion correction. \href{https://doi.org/10.1002/jcc.20495}{\textit{J. Comput. Chem.} \textbf{27,} 1787 (2006)}.

\bibitem{SIMagonov}  Magonov, S. N. et al. Scanning tunneling and atomic force microscopy study of layered transition metal halides ${\mathrm{Nb}}_{3}{\mathrm{X}}_{8}$ ($\mathrm{X}$ = $\mathrm{Cl}$, $\mathrm{Br}$, $\mathrm{I}$).  \href{https://doi.org/10.1021/ja00059a053}{\textit{J. Am. Chem. Soc.} \textbf{115,} 2495 (1993)}.

\bibitem{SIJiangNb3X8} Jiang, J. et al. Exploration of new ferromagnetic, semiconducting and biocompatible ${\mathrm{Nb}}_{3}{\mathrm{X}}_{8}$ ($\mathrm{X}$ = $\mathrm{Cl}$, $\mathrm{Br}$ or $\mathrm{I}$) monolayers with considerable visible and infrared light absorption. \href{https://doi.org/10.1039/C6NR07231C}{\textit{Nanoscale} \textbf{9,} 2992 (2017)}.

\bibitem{SIKresse1996} Kresse, G. \& Furthmüller, J. Efficient iterative schemes for ab initio total-energy calculations using a plane-wave basis set. \href{https://doi.org/10.1103/PhysRevB.54.11169}{\textit{Phys. Rev. B} \textbf{54,} 11169 (1996)}.

\end{thebibliography}
\end{document}